%% file: LoraDriftPaper.tex
\documentclass[conference, a4paper]{IEEEtran}

\input{IEEEincludes}

\usepackage{amsmath}

\usepackage[detect-weight=true, binary-units=true]{siunitx} 
\usepackage[nolist,nohyperlinks]{acronym} 
\usepackage{tikz}
\usepackage{pgfplots}
\usepackage{dsfont}					
\usepackage{amssymb}				

\usepackage[caption=false,font=footnotesize]{subfig} 
\usepackage{paralist}			
\usepackage{placeins}
\usepackage{graphicx}
\usepackage{multirow}
\usepackage{xcolor}
\usepackage{soul}

\pgfplotsset{compat=newest}
\usepackage{pgfplotstable}

 \graphicspath{{figures/}{figures/svg/}}

\usepackage[xindy,acronyms,nopostdot]{glossaries}

\usetikzlibrary{plotmarks}
\usepackage{circuitikz}
\usetikzlibrary{calc,positioning,shapes,decorations.pathreplacing,arrows,arrows.meta,circuits,circuits.ee.IEC}

\input{commands}

\newlength\figureheight
\newlength\figurewidth
\newlength\figureheightW
\newlength\figurewidthW
\newlength\figureheightN
\newlength\figurewidthN
\begin{document}
%
\input{acronyms.tex}

\title{
Impact of Phase Noise and Oscillator Stability on Ultra-Narrow-Band-IoT Waveforms for Satellite }
%
%

\author{\IEEEauthorblockA{{Christian A. Hofmann}, {Kai-Uwe Storek} and {Andreas Knopp}}
\IEEEauthorblockA{Bundeswehr University Munich, Chair of Signal Processing, 85579 Neubiberg, Germany.\\
Email: papers.sp@unibw.de}
}%

\thanks{Manuscript received xxxx xx, 20xx; revised xxxx xx, 20xx.
accepted
xxxx xx, 20xx. Date of publication xxxx xx, 20xx; 
The authors are with the Signal Processing Group of the Institute of Information Technology, Bundeswehr University Munich, 85579 Neubiberg,
Germany (e-mail:  papers.sp@unibw.de).}%

\newif\ifhlght
\hlghttrue 

\newcommand{\revcolor}{blue}
\newcommand{\revcolorB}{black}
\newcommand{\revcolorC}{red}

\markboth{ }%
{}
%



\maketitle

\begin{abstract}
It has been shown that \gls{unb} massive machine type communication using very compact devices with direct access to satellites is possible at ultra low rate. This enables  global ubiquitous coverage for terminals without terrestrial service in the Internet of Remote Things and provides access to any satellite up to the the geostationary
earth orbit. 
The lower data rate for waveforms providing \gls{unb} communication is set by the stability and the phase noise of the applied oscillators. 
In this paper we analyze the physical layer of two candidate waveforms, which are LoRa and \gls{ucss} with respect to phase noise and oscillator frequency drifts. It is figured out that \gls{ucss} is more robust against linear frequency drifts, which is the main source of error for \gls{unb} transmissions. 
\end{abstract}

%
%

%
\IEEEpeerreviewmaketitle

\section{Introduction}
\label{sec:intro}
\glsresetall
\Gls{nb} \gls{mmtc} will be an indispensable part of future mobile communication. For worldwide, seamless communication or continued service in remote areas, satellite systems enable global ubiquitous coverage in the \gls{iort} 
\cite{DeSanctis2016}. 
Satellite constellations in the \gls{leo} provide such coverage with a large number of satellites. They benefit from lower path loss and shorter latency compared to satellites at \gls{meo} or \gls{geo}. 
The large number of required spacecraft for a worldwide, uninterrupted service leads to high costs for the system and subsequently high costs per transmitted bit for the subscribers.  
Satellite systems in \gls{geo} enabling \gls{mtc} in the \gls{iot} outside the polar regions with a small number of satellites usually operating at frequencies of the L-band or S-band. 
As these frequency bands do not offer much bandwidth, the costs for providing communication services are also quite high. 
At higher frequency bands like the C-Band or above, the link budget to \gls{geo} satellites offers only an \gls{unb} communication with a few bits per second if very small transceivers are assumed and regulatory limits are met. Nevertheless, the large available bandwidths allow \gls{unb} communication for a vast number of devices at low cost. In 
\cite{Hofmann2019}
a novel \gls{unb} waveform called \gls{ucss} has been presented that enables the random multiple access of very small devices to satellites in the \gls{geo}. The feasibility has been demonstrated at X-Band and Ku-Band frequencies\cite{Hofmann2019a}. 
The key to \gls{unb} communication is the efficient acquisition and synchronization of the signal that is received at very low power. 
Usually, the acquisition and tracking performance of the applied algorithms sets the lower limit for the data rate. 
It is not simply possible to operate a waveform at a lower symbol rate to reduce the data rate and close the link budget. 
It has been shown in \cite{Hofmann2019a} that phase noise is limiting the correlation gain of spread spectrum waveforms and, hence, sets a lower limit for the symbol rate. 
In this paper we analyze the impact of phase noise on \gls{unb} communication in detail and demonstrate that the components of the phase noise with lower frequencies confine the symbol rate.  We consider \gls{ucss} and LoRa, a \gls{css} waveform, as candidates for uNB waveforms for satellite direct access.

In Section \ref{sec:css}, LoRa and \gls{ucss} are presented and the expected impact of phase noise on the waveforms is analyzed theoretically. 
Section \ref{sec:sim} presents simulation results for the performance of the investigated waveforms under the impact of phase noise and linear frequency drifts.
In Section 
\ref{sec:meas},
we provide measurement results that confirm the theoretical analysis and the simulation results. 
We conclude the paper in Section
\ref{sec:concl}.
 
\section{Chirp Spread Spectrum Narrowband Waveforms }
\label{sec:css}
\subsection{LoRa Basics}
LoRa is a wireless communication technology that applies \gls{css} for the modulation and enables communication at very low rate for terrestrial services \cite{Raza2017,Mroue2018}.
It is also considered for applications in \gls{leo} satellite systems \cite{Lacuna2020}.  
LoRa uses a \gls{mfsk} modulation that is spread by linear chirps. An important parameter is the spreading factor $\SF$, which is usually given by $\SF = 2^\loraSF$. Each symbol carries $\loraSF$ bits and, hence, LoRa is a $\SF$-ary \gls{mfsk} with $\SF$ tones and chirp spreading.
 
For the transmission of $\PL$ bytes as payload, LoRa uses $\loraSmbNr$ symbols if no payload header is added \cite{SemtechCorporation2013}. 
\begin{equation}
\loraSmbNr= 8 + \text{max}\left\{  \left\lceil \frac{8\loraPL - 4 \cdot \loraSF +24}{4\left( \loraSF - 2 \cdot \loraLDRon \right)} \right\rceil \left(\loraCR + 4 \right),0 \right\}
\label{eq:1}
\end{equation}
Here, 
$\loraCR = 1,...,4$ is used for the calculation of the coding rate $\CR=\frac{4}{4+\loraCR}$, and
$\loraLDRon$ indicates the enabled \gls{ldro}.
$\loraLDRon=1$ stands for enabled , $\loraLDRon=0$ for disabled).  The \gls{ldro} parameter is usually set when the LoRa symbol time is equal or above 16.38 ms \cite{Semtech2017}.

The time $\TS$ for the transmission of one chirp, i.e. one symbol, is given by
$\TS = \frac{\SF}{\BW} $, where $\BW=1/\sampleTime$ stands for the sample rate. 

Thus, all symbols of one frame are transmitted within the time  
$\loraTpay =\TS \cdot \loraSmbNr $.

\begin{figure}[t]
\footnotesize
	\centering
	\scalebox{0.9}{
	\input{figures/TikzDraw/overhead/source.tex}
}
	\caption{LoRa and UCSS frame structure}
	\label{fig:packet}
\end{figure}
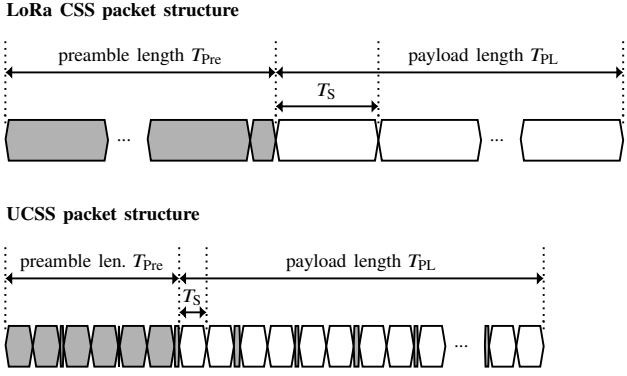

LoRa typically uses $\PreSymbs = 12.25$ symbols for the preamble, which then covers a time   
$\loraTpre = \TS \cdot \loraPreSymbs$. The structure of a transmitted LoRa frame is depicted in Fig.\,\ref{fig:packet}. 
Finally, the data rate in bit/s is given by
\begin{equation}
	\loraDR = \frac{8 \cdot\PL}{\loraTpre+\loraTpay}
\end{equation}
and the transmission overhead $\OHLora$ is 
\begin{equation}
	\OHLora = 1 - \frac{\loraTpay}{\loraTpac}.
\end{equation}
Table
\ref{tab:lora}
lists the calculated parameters for different settings of $\BW$ and $\loraSF$, which are also used in this paper for simulations and measurements. 


\subsection{Impact of Phase Noise on LoRa}
\label{sec:loraPN}
It has been shown that the requirements on the phase noise performance of oscillators are quite relaxed when \gls{fsk} modulation is used
\cite{Chen2017a}. As also shown by simulation in subsection \ref{sec:pn}, the faster phase noise components (i.e. the white frequency-, flicker phase-, and white phase noise component) are not harmful to LoRa modulation. On the the other hand, random frequency walk and flicker frequency noise are affecting its performance. 
For LoRa, the frequency drift of the \gls{lo} is one of the most relevant parameters limiting the maximum packet length \cite{SemtechCorporation2018}.
This becomes obvious if the \gls{cfo} between \gls{tx} and \gls{rx} is considered. If we assume that the \gls{cfo} is estimated and compensated perfectly only at the beginning of a packet, then a symbol error occurs whenever a tone out of the \gls{mfsk} alphabet is shifted towards the neighboring tone beyond the decision boundary.
With LoRa, the distance in frequency between any two tones is given by \cite{Semtech2017}
\begin{equation}
\FreqDiff = \frac{\BW}{\SF}.
\label{eq:loraFdist}
\end{equation}

This means that a symbol error occurs if the \gls{cfo} is larger than $0.5 \cdot \FreqDiff$. This leads to the limit for the maximum frequency offset $\loraFreqDriftMaxDS$ that may occur during the timespan of one frame 
\begin{equation}
\loraFreqDriftMaxDS = \frac{\BW}{3\cdot 2^\loraSF},
\label{eq:maxDriftLoraDS}
\end{equation}
provided in the LoRa documentation \cite{Semtech2017}.

For higher values of $\loraSF$ and lower values of $\BW$, the neighboring symbols have a quite small distance $\FreqDiff$, and the modulation is very sensitive to drifting CFO.
Therefore, the \gls{ldro} parameter is set ($\loraLDRon=1$) if $\TS \geq 16.38$ ms.
In that case, the number of bits per symbol is reduced by two
and the maximal frequency shift during a single packet transmission provided in \cite{Semtech2017} is relaxed to $\loraFreqDriftMaxDSLDRO$  given by 
\begin{equation}
\loraFreqDriftMaxDSLDRO = 16 \cdot \loraFreqDriftMaxDS.
\label{eq:maxDriftLoraCal}
\end{equation} 

This value given by the data sheet is larger than the frequency difference between neighboring tones with $\loraLDRon=1$ given by
\begin{equation}
\FreqDiff = 	\frac{\BW}{2^{\loraSF-2}}.
\end{equation}
This means that practical LoRa receivers somehow track the CFO to improve the resistance against frequency drift.   
If we assume a linear drift of the CFO with rate $\freqDriftRate$, then the maximum drift rate 
$\maxfreqDriftRate$ defining the upper limit for the successful transmission of a frame is calculated from the length of the payload $\Tpay$ by%
\footnote{Here, we assume that the CFO is compensated perfectly at the end of the preamble.}
\begin{equation}
 \maxfreqDriftRate = \frac{\FreqDriftMax}{\Tpay}.	
\end{equation}

For LoRa devices an analysis on CFO drift of practical devices is published in 
\cite{SemtechCorporation2018}, where an almost linear frequency drift over time is observed, while the circuit is heated by the \gls{rf} components in burst transmission. Drift rates $\freqDriftRate = \frac{\delta f(t)}{\delta t}$ between \SI{20}{Hz/s} and \SI{80}{Hz/s} are reported for a Semtech reference design PCB at \SI{915}{MHz}, at $25^\circ$C in a +\SI{15}{dBm} transmit operation.
The drift is reduced to values between \SI{10}{Hz/s} and \SI{30}{Hz/s} by thermal isolation of the oscillator and to 
\SI{< 20}{Hz/s} if a \gls{tcxo} is used.

In \cite{Doroshkin2019} LoRa has been analyzed with respect to dynamic Doppler shifts and limits for maximum frequency drifts are determined by measurements.
The authors delivered, the maximum allowable frequency drift of \SI{4.94}{ppm/s} and \SI{0.33}{ppm/s} at \SI{434}{MHz} with symbol rates of \SI{250}{kHz} and \SI{125}{kHz} for SF 11 and 12, respectively. These values are higher than than the specified values in \cite{Semtech2017} and couldn't be confirmed by our measurements.

From Table \ref{tab:lora} it is derived that the drift rates provided in \cite{SemtechCorporation2018} are too high to guarantee a successful transmission with all settings for $\BW$ and $\loraSF$. 

A countermeasure for a transmission at the given drift rates $\freqDriftRate$ at lower symbol rate is to shorten the frame length. By this, $\Tpay$ becomes shorter and $\maxfreqDriftRate$ is increased. Unfortunately, the LoRa transmission is very inefficient at very short frame lengths, 
because the fixed preamble length of the standard produces a significant amount of overhead.
Therefore, we consider \gls{ucss} in the following, which has been developed with respect to efficient transmission with short frames.

\setlength\tabcolsep{2.5pt} 
\renewcommand{\arraystretch}{1.5}
\begin{table*}
\scriptsize
    \centering
    \caption{Settings and parameters for LoRa and UCSS. Equal setting numerals for both waveforms indicate similar required SNR for successful transmission}
\input{tabledata}
	\label{tab:lora}
\end{table*}

\subsection{UCSS Basics}
\label{sec:ucss}

\gls{ucss} promises efficient communication with reduced overhead at very low rates \cite{Hofmann2019b}. The main idea is to introduce short pause times between the chirps to separate and identify users transmitting simultaneously, while the phase of the chirps is modulated to carry information. With \gls{ucss} it is possible to transmit at ultra-low rate and provide access to satellites in GEO for very small devices at higher frequencies, where sufficient spectrum is available for \gls{mmtc}.

For \gls{ucss}, only a short header with pilot symbols for \gls {cfo} estimation is required as depicted in Fig.\,\ref{fig:packet}. Further overhead is caused by the pause times between the symbols, which become longer, the more symbols are transmitted within one packet. 
Hence \gls{ucss} is suitable for efficient transmission of short blocks at ultra-low \glspl{snr}.

We provide settings for the spreading factor $\SF$ and the symbol rate $\BW$ of UCSS that require similar sensitivity like the settings for LoRa. Equal setting numerals in Table \ref{tab:lora} for both waveforms indicate similar required SNR for successful transmission. Here, the waveforms are compared by  their  \gls{fer} over the \gls{snr} measured by simulations in \gls{awgn} without further errors or synchronization. The values for the required SNR in Table \ref{tab:lora} are found in \cite{Semtech2017}.   

Fairness is ensured as the waveforms are compared when operating in the same bandwidth with the same transmit power. Although UCSS is a waveform enabling random multiple access, we consider only one active user here, for the sake of a fair comparison.      

The number of symbols $\SmbNr$ within a UCSS frame is calculated by 
\begin{equation}
  	\ucssSmbNr = \frac{\PL\cdot8+\CRC}{\CR}
  \end{equation}  
  where $\CRC = 3$ is the number of bits used for a \gls{crc}.
Here, we use a header with $\PreSymbs = 6$ symbols for \gls{cfo} estimation. 
The length $\Tpause$ of the pause times between the symbols depends on the used unipolar code \cite{Hofmann2019b} and is given by 
\begin{equation}
	\Tpause = \left\lceil\ucssSmbNr /2 \right\rceil^2 / \BW.
\end{equation}
Thus, the duration of the entire frame is given by 
\begin{equation}
	\ucssTpac = \ucssTpre + \ucssTpay + \Tpause,
\end{equation}
and the transmission overhead $\OHUcss$ is 
\begin{equation}
	\OHUcss =1- \frac{\ucssTpay}{\ucssTpac}.
\end{equation}


\subsection{Impact of Phase Noise on UCSS}
As UCSS is a differentially modulated waveform, a symbol error due to a CFO is very unlikely. In fact, a false decision for a symbol occurs if the CFO shifts the phase during the time between two \gls{dbpsk} symbols by more than $\pi/4$. The resulting values $\ucssFreqDriftMax$ for the CFO causing a symbol error are quite high. Instead, the maximum CFO is limited for UCSS by the timeshift of the chirp symbols. As any frequency shift of a chirp is linked to a shift of the chirp in time domain, the chirp is shifted by the CFO in time and no longer detected if the CFO is too high. 
Thus, the maximum CFO is not given by $\ucssFreqDriftMax$ but is instead approximately calculated by
\begin{equation}
	\ucssFreqDriftMax \approx \frac{1}{2} \cdot \frac{\BW}{\SF}.
\end{equation}
This is further analyzed by simulation in Section \ref{sec:drift}.

\section{Simulation Results }
\label{sec:sim}

\subsection{Phase Noise }
\label{sec:pn}

\begin{figure}[t]
\footnotesize
	\centering
	\input{figures/pnProfiles.tex}
	\caption{Phase noise power spectral density (PSD) profiles used for simulation of phase noise}
	\label{fig:pnProfiles}
\end{figure}
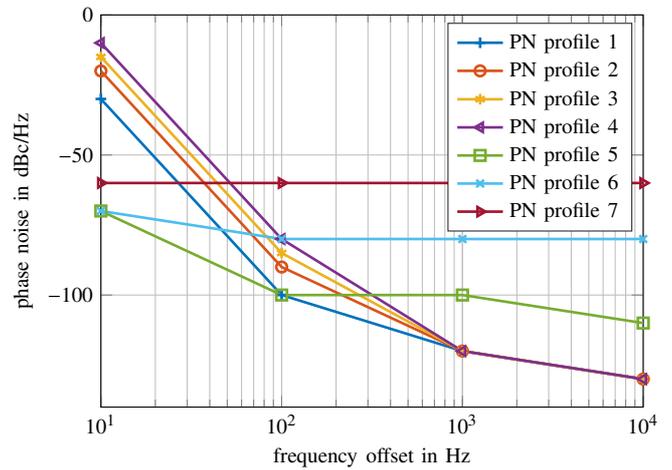
\begin{figure*}[t]
\footnotesize
	 \centering
	 
     \subfloat[LoRa frame error rate; $\BW$ = \SI{20}{kHz}, $\loraSF$ = 11 (setting LS-4)]{ \centering
        \input{figures/LoraPnSimFer.tex}%
        }%
	 \label{fig:LoraPnSimFer}%
    %
~~~~
\subfloat[UCSS frame error rate; $\BW$ = \SI{20}{kHz}, $\SF$ = 211 (setting US-4).]{ \centering
	\input{figures/UcssPnSimFer.tex}
	}
	 \label{fig:UcssPnSimFer}
	 \caption{Frame error rate over the SNR for different phase noise PSD profiles as given in Fig.\,\ref{fig:pnProfiles}}
	  \label{fig:PnSimFer}%
\end{figure*}
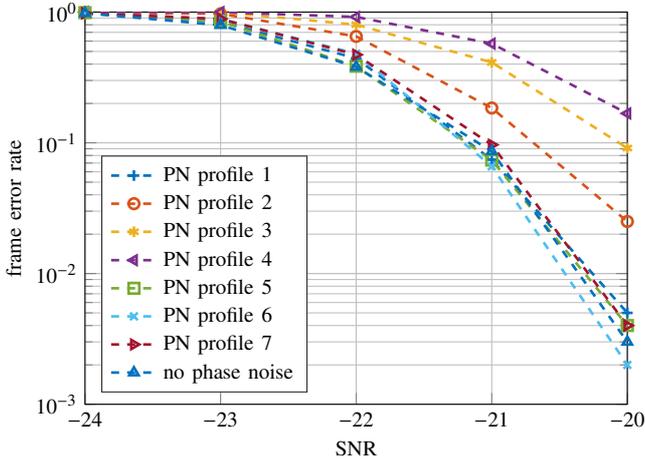
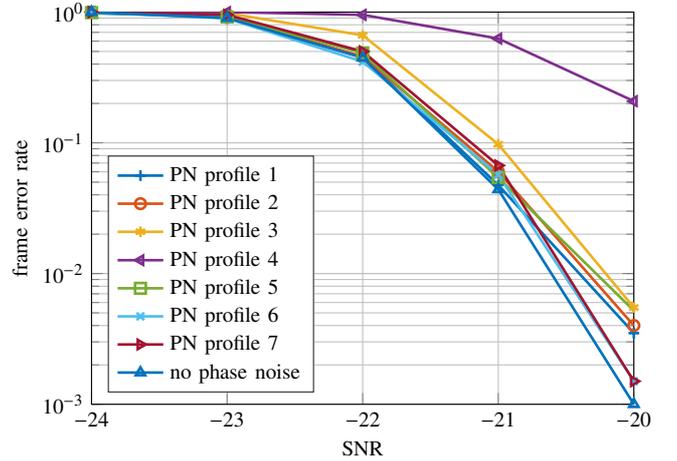

In this subsection we analyze the impact of phase noise on the performance of LoRa and UCSS. First, we demonstrate that both CSS waveforms perform well under the influence of fast phase noise (white frequency-, flicker phase-, and white phase noise), while slow phase noise (random frequency walk and flicker frequency) is harmful to UCSS and LoRa. 
Therefore, we use different phase noise 
\gls{psd} 
profiles as depicted in 
Fig.\,\ref{fig:pnProfiles}.
Starting at the realistic \mbox{profile 1}, we increase only the slow phase noise from \mbox{profile 1} to \mbox{profile 4}. From profile 5 to profile 7, the fast phase noise is increased incrementally. 

The simulation is performed in MATLAB using the phase noise system object, which applies a filtered, white noise signal for the phase error. The filter coefficients are calculated from the profiles of Fig.\,\ref{fig:pnProfiles} according to \cite{Kasdin1995a}.

For the simulation of LoRa we applied the code from \cite{Mroue2018} and further implemented the \gls{ldro}. The simulations were performed in \gls{awgn} without introducing further errors (frequency, clock) and further synchronization algorithms at the receiver.
The results for the \gls{fer} over the \gls{snr} are shown in Fig.\,\ref{fig:PnSimFer}a. The figure reveals that only profiles 2, 3 and 4 are negatively affecting the performance of LoRa, indicated by an increased \gls{fer}.

We repeated the simulation using UCSS with the parameters provided in the caption of Fig.\,\ref{fig:PnSimFer}b. The presented results there also indicate, that only the slow phase noise degrades the performance of UCSS. 
In comparison to LoRa, it can be observed that LoRa is slightly stronger affected by slow phase noise components%
. 

We focus on slow phase noise in the remainder of this paper as this is the most harmful distortion for UCSS and LoRa. We consider the linear frequency drift as a kind of worst case and, thus, we determine the maximum CFO drift rate for both waveforms.

\subsection{Linear Frequency Drift}
\label{sec:drift}

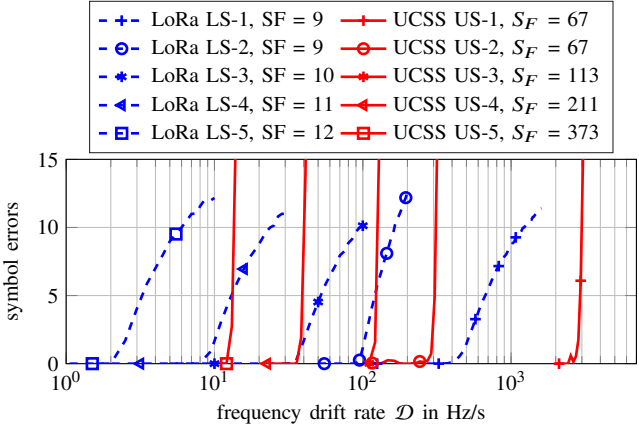
\begin{figure}[t]
\footnotesize
	\centering
	\input{figures/driftSimSymErrs.tex}
	\caption{Number of symbol errors for LoRa and UCSS versus the the frequency drift rate. Marker symbols indicate  the same requirement on the SNR. }
	\label{fig:driftSimSymErrs}
\end{figure}
\begin{figure}[t]
\footnotesize
	\centering
	\input{figures/driftSimFer.tex}
	\caption{Simulation results for the frame error rate (FER) of LoRa and UCSS versus the the frequency drift rate. Marker symbols indicate  the same requirement on the SNR.}
	\label{fig:driftSimFer}
\end{figure}
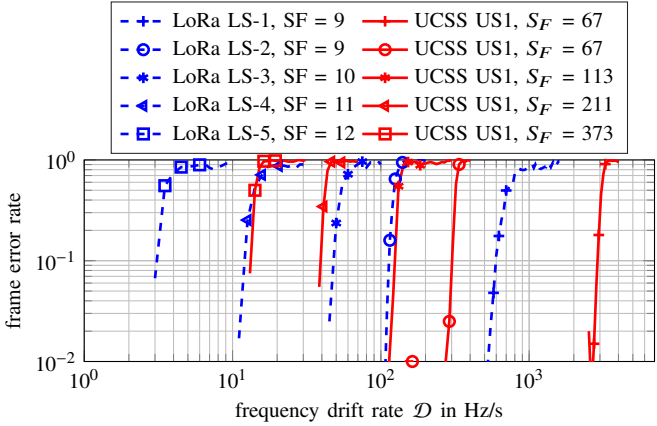

In the following, we present simulation results for LoRa and UCSS under the influence of a linearly increasing CFO. Here, the $\txSigLen$ receive signal samples $\rxSig[\bin]$ with $n = 0,...,\txSigLen-1$ are calculated from the  transmit signal samples $\txSig[\bin]$
by
\begin{equation}
	\rxSig[\bin] = \txSig[\bin] \cdot \phi \cdot \Phi[n]+  \noiseSig[\bin] 
\end{equation}
where $\phi$ is a random phase in $[0,2\pi[$ and $\noiseSig[\bin]$ stands for complex Gaussian noise with variance $\sigma_\eta^2$ so that the desired SNR $\snr=\frac{\sigma_x^2}{\sigma_\eta^2}$ is achieved. The CFO drift $\Phi[n]$ is 
\begin{equation}
	\Phi[n] = \myexp{}{2\pi\left( \bin \frac{\freqDriftRate \cdot \Tpay}{\txSigLen} \right) \bin \sampleTime}
\end{equation}
to ensure that the CFO $\CFO$ after $\Tpay$ is $\CFO = \freqDriftRate\cdot \Tpay$.

The Monte Carlo simulations in \gls{awgn} are performed at an SNR ensuring a   \SI{3}{dB} margin from the required SNR provided in Table \ref{tab:lora}.  
When regarding Fig.\,\ref{fig:driftSimSymErrs}, it is observed that the number of symbol errors is increasing over the increasing frequency drift rate $\freqDriftRate$.
In the figure, the blue dashed curves stand for LoRa and the red solid curves for UCSS. The same marker symbols indicate comparable settings for both modulations with the same requirement on the SNR. The figure reveals that UCSS is more robust against drifting CFO than LoRa for all settings analyzed here. This is also derived from    
Fig.\,\ref{fig:driftSimFer} 
where the \gls{fer} is plotted over the frequency drift rate. 
%

\section{Measurement Results }
\label{sec:meas}

\begin{figure}[t]
\footnotesize
	\centering
	\scalebox{0.9}{
	\input{figures/TikzDraw/measSetup/source.tex}
}
	\caption{Measurement Setup}
	\label{fig:measSetup}
\end{figure}
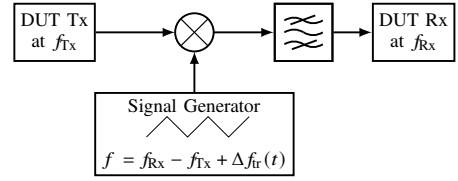

To verify the simulation results in practice, we performed measurements with LoRa hardware and a \gls{sdr} implementation of UCSS.
We used two 
Semtech SX1261/SX1262 Development Kits as LoRa \gls{tx} and \gls{rx}. For UCSS, 
an 
SDR Ettus USRP B205 was used as \gls{tx} and a B210 as \gls{rx}. 
The devices were connected in a test setup according to Fig.\,\ref{fig:measSetup}, where the Tx was operated at center frequency $\fTx$ while the receiver was receiving at $\fRx$. A frequency conversion with a time variant \gls{lo} $\CFOtr(t)$ introduced a linear frequency drift. Here, $\CFOtr(t)$ was a triangle wave with
\begin{equation}
	\CFOtr(t)=2\cdot \CFOmeasMax \left|\frac{t}{\periodTri}-\left\lfloor \frac{t}{\periodTri}+\frac{1}{2}\right\rfloor \right|,
\end{equation}
where
$\periodTri$ is the period of the triangle wave given by 
\begin{equation}
	\periodTri=\frac{\CFOmeasMax}{\freqDriftRateMeas}
\end{equation}
to ensure a linear frequency drift with rate $\freqDriftRateMeas$.
The maximum CFO $\CFOmeasMax$ was chosen such that the receivers were still able to operate at $\CFOmeasMax$ without performance degradation. We used $\CFOmeasMax= \frac{1}{4}\BW$ in our tests. During the measurements we used $\fTx$ = \SI{830}{MHz} and $\fRx$ = \SI{880}{MHz} for both waveforms.   

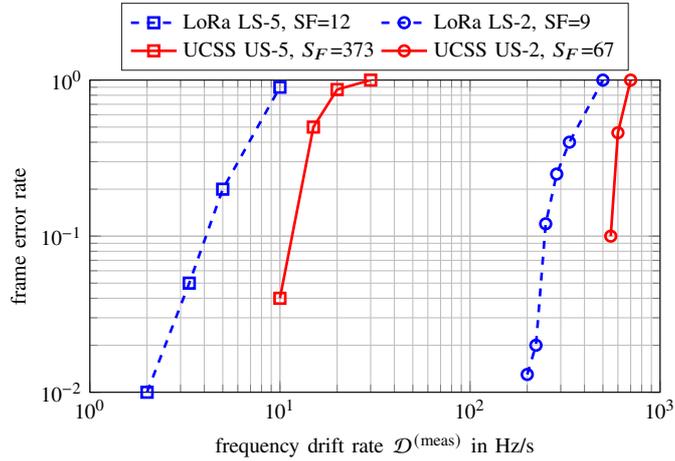
\begin{figure}[t]
\footnotesize
	\centering
	\input{figures/combDriftMeas.tex}
	\caption{Measured frame error rate (FER) for LoRa and UCSS at \mbox{$\fTx$ = \SI{830}{MHz}} and \mbox{$\fRx$ = \SI{880}{MHz}} with $\BW$ = \SI{20}{kHz} }
	\label{fig:combDriftMeas}
\end{figure}

Fig.\,\ref{fig:combDriftMeas}
depicts the measured \gls{fer} for LoRa and UCSS over the frequency drift rate $\freqDriftRateMeas$. 
The measurement results confirm the simulation very well. For LoRa it is observed that the robustness against CFO drift predicted by the datasheet is confirmed in practice for higher data rates. 
At the ultra low rate given by setting LS-5, the datasheet predicts a maximum rate of \SI{4.5}{Hz/s}, while only \SI{2}{Hz/s} were tolerable in the measurements.
The setting with the lowest data rate for UCSS (US-5) was able to withstand a CFO drift of up to \SI{10}{Hz/s} in our measurements. 
This is approximately \SI{5}{Hz/s} less than observed from the simulation results. 
Of course the frequency drift of the oscillators of Tx and Rx hardware during the measurements cannot be ignored and is added to the drift introduced by the LO. Although we kept the Tx and Rx in an environment with constant temperature and operated them in continuous mode to avoid heating of the LO in burst mode, the residual drift obviously affected the measured results. 
Nevertheless, the measurement results are very close to the simulation results and finally confirm our theoretical assumptions.         

Both waveforms can be adopted to withstand CFO drifts with even higher rates than presented here if they occur in practice. This is achieved by shorter frames with less bytes within the payload. Another countermeasure to CFO drifts would be the insertion of pilot symbols within the frame, which of course also increases the overhead. The relationship between payload length and robustness against CFO drift is shown in Table \ref{tab:lora} by the comparison of the \mbox{settings 5} \mbox{and 6}.  
The drawback for LoRa is the increased overhead if the payload is shortened. This is derived from Table \ref{tab:lora} where the communication overhead for LoRa is increasing from 43\% to 51\% for \SI{8}{bytes} and \SI{4}{bytes} per frame in setting LS-5 and LS-6, respectively. 
On the other hand, the overhead of UCSS is slightly reduced if shorter frames are used in settings US-5 and US-6. Hence, UCSS not only shows a better robustness against CFO drifts than LoRa but also is able to transmit with less overhead.  


\section{Conclusion}
\label{sec:concl}
In this paper we analyzed the impact of phase noise and frequency drift on two candidate waveforms for satellite ultra-narrow-band transmission. With LoRa and UCSS, we considered and compared two CSS waveforms by simulation and measurements in the laboratory. 
Both waveforms are not sensitive to fast phase noise components, while the slower components of the phase noise (random frequency walk and flicker frequency noise) are affecting the transmissions. 
We concentrated our analysis to the most harmful phase error, which is a linear frequency drift. 
We derived the maximum tolerable drift rate for the successful transmission of a frame at different sample rates and spreading factors. 
UCSS showed an increased robustness against drifting carrier frequencies in all investigated settings  compared to the equivalent LoRa setting. Furthermore, UCSS allows the shortening of the transmitted frames without increasing the transmission overhead and, thus, is able to increase the robustness against linear frequency drifts if necessary. 

\ifCLASSOPTIONcaptionsoff
  \newpage
\fi




\bibliographystyle{myIEEEtran}
\bibliography{IEEEabrv,bib/library}
\end{document}

%% file: IEEEincludes.tex
\usepackage[T1]{fontenc}

\usepackage{cite}

\ifCLASSINFOpdf
\else
\fi
\usepackage{amsmath}
\usepackage[cmintegrals]{newtxmath}

%% file: commands.tex
\newcommand{\loraSmbNr}{N_\text{S}^\text{(LoRa)}}
\newcommand{\loraPL}{N_\text{byte}^\text{(LoRa)}}

\newcommand{\loraSF}{\text{SF}}
\newcommand{\loraLDRon}{\text{LDR}}
\newcommand{\loraCR}{\text{CR}}

\newcommand{\loraFreqDriftMaxDS}{\Delta f_\text{max}^{(\text{LoRa})}}
\newcommand{\loraFreqDriftMaxDSLDRO}{\Delta f_\text{max}^{(\text{LoRa,LDR})}}

\newcommand{\loraTpay}{T_\text{PL}^\text{(LoRa)}}
\newcommand{\loraPreSymbs}{N_\text{pre}^\text{(LoRa)}}
\newcommand{\loraTpre}{T_\text{pre}^\text{(LoRa)}}
\newcommand{\loraTpac}{T_\text{fr}^\text{(LoRa)}}
\newcommand{\loraDR}{R_\text{bit}^\text{(LoRa)}}

\newcommand{\ucssSmbNr}{N_\text{S}^\text{(UCSS)}}

\newcommand{\ucssFreqDriftMax}{\Delta f_\text{max}^\text{(UCSS)}}

\newcommand{\ucssTpay}{T_\text{PL}^\text{(UCSS)}}

\newcommand{\ucssTpre}{T_\text{pre}^\text{(UCSS)}}
\newcommand{\ucssTpac}{T_\text{fr}^\text{(UCSS)}}

\newcommand{\FreqDriftMaxDS}{\Delta f_\text{max}}

\newcommand{\SF}{{S_F}}
\newcommand{\CR}{R_\text{c}}
\newcommand{\PL}{N_\text{byte}}
\newcommand{\FreqDriftMax}{\Delta f_\text{max}}
\newcommand{\FreqDiff}{\Delta f_\text{S}}

\newcommand{\BW}{R_\text{s}}
\newcommand{\sampleTime}{T_\text{s}}
\newcommand{\SmbNr}{N_\text{S}}
\newcommand{\CRC}{N_\text{CRC}}
\newcommand{\TS}{T_\text{S}}
\newcommand{\Tpay}{T_\text{PL}}
\newcommand{\PreSymbs}{N_\text{Pre}}
\newcommand{\Tpre}{T_\text{Pre}}
\newcommand{\Tpac}{T_\text{Fr}}
\newcommand{\Tpause}{T_\text{Pause}}
\newcommand{\DR}{R_\text{bit}}

 \newcommand{\carrierFreq}{f}
 \newcommand{\CFO}{\Delta \carrierFreq}

\newcommand{\freqDriftRate}{\mathcal{D}}
\newcommand{\maxfreqDriftRate}{\mathcal{D}_\text{max}}

\newcommand{\freqDriftRateMeas}{\mathcal{D}^{(\text{meas})}}
\newcommand{\OHLora}{O^\text{(LoRa)}}
\newcommand{\OHUcss}{O^\text{(UCSS)}}
\newcommand{\txSig}{x}

\newcommand{\bin}{n}
\newcommand{\txSigLen}{N_\text{rx}}
\newcommand{\fTx}{f_\text{Tx}}
\newcommand{\fRx}{f_\text{Rx}}
\newcommand{\CFOtr}{\Delta \carrierFreq_\text{tr}}
\newcommand{\CFOmeasMax}{\Delta f_\text{max,tr}}
\newcommand{\periodTri}{T_\text{tr}}
\newcommand{\snr}{\Gamma}
\newcommand{\snrReq}{\snr_\text{min}}

 \newcommand{\noiseSig}{{\eta}}

\newcommand{\rxSig}{r}

\newcommand{\imj}{\operatorname{j}}

\newcommand{\myexp}[2]{{\exp\left\{#1\imj #2\right\}}}

%% file: acronyms.tex
\setacronymstyle{long-short}
\newacronym{af}{AF}{amplify-and-forward}
\newacronym[description={naive \gls{af}},user1={Naive Amplify-and-Forward}]{naf}{NAF}{naive amplify-and-forward}
\newacronym[description={advanced \gls{af}},user1={Advanced Amplify-and-Forward}]{aaf}{AAF}{advanced amplify-and-forward}
\newacronym{siso}{SISO}{single-input single-output}
\newacronym{mimo}{MIMO}{multiple-input multiple-output}
\newacronym[description={multiuser \gls{mimo}}]{mu-mimo}{MU-MIMO}{multiuser multiple-input multiple-output}
\newacronym[user1={Minimum Mean Square Error}]{mmse}{MMSE}{minimum mean square error}
\newacronym{zf}{ZF}{zero-forcing}
\newacronym{hts}{HTS}{high throughput satellite}
\newacronym{mrc}{MRC}{maximum ratio combining} 
\newacronym{cnir}{CNIR}{carrier to noise plus interference ratio} 
\newacronym{los}{LOS}{line-of-sight}
\newacronym{dpc}{DPC}{dirty paper coding}
\newacronym{mse}{MSE}{mean squared error}
\newacronym{qos}{QoS}{quality-of-service}
\newacronym{hpa}{HPA}{high power amplifier}
\newacronym{svd}{SVD}{singular value decomposition}
\newacronym{dtp}{DTP}{digital transparent processing}
\newacronym{obp}{OBP}{on-board processing}
\newacronym{awgn}{AWGN}{additive white Gaussian noise} 
\newacronym{ecef}{ECEF}{earth centered, earth fixed}
\newacronym{psk}{PSK}{phase-shift keying}
\newacronym{fsk}{FSK}{frequency-shift keying}
\newacronym[description={multiple \gls{fsk}}]{mfsk}{MFSK}{multiple frequency-shift keying}
\newacronym{qam}{QAM}{quadrature amplitude modulation}
\newacronym{eirp}{EIRP}{effective isotropic radiated power}
\newacronym{kkt}{KKT}{Karush-Kuhn-Tucker}
\newacronym{thp}{THP}{Tomlinson-Harashima precoding} 
\newacronym[description={quadrature \gls{psk}}]{qpsk}{QPSK}{quadrature phase-shift keying}  
\newacronym{pdf}{PDF}{probability density function} 
\newacronym{csi}{CSI}{channel state information} 
\newacronym{rf}{RF}{radio frequency} 
\newacronym{ml}{ML}{maximum likelihood}
\newacronym{blue}{BLUE}{best linear unbiased estimator}  
\newacronym{tdma}{TDMA}{time division multiple access} 
\newacronym{fdma}{FDMA}{frequency division multiple access} 
\newacronym{cdma}{CDMA}{code division multiple access} 
\newacronym{scma}{SCMA}{scrambled coded multiple access} 
\newacronym{idma}{IDMA}{interleave division multiple access} 
\newacronym[shortplural={FSS},firstplural={fixed satellite services (FSS)}]{fss}{FSS}{fixed satellite service}
\newacronym{nasa}{NASA}{national aeronautics and space administration}
\newacronym{dsn}{DSN}{deep space network}
\newacronym{vod}{VoD}{Video-on-Demand} 
\newacronym{ctm}{CTM}{channel transfer matrix} 
\newacronym{tx}{Tx}{transmitter}
\newacronym{rx}{Rx}{receiver} 
\newacronym{obo}{OBO}{output backoff} 
\newacronym{satcom}{SATCOM}{satellite communication} 
\newacronym{ula}{ULA}{uniform linear array} 
\newacronym{geo}{GEO}{geostationary earth orbit} 
\newacronym{meo}{MEO}{medium earth orbit} 
\newacronym{leo}{LEO}{low earth orbit} 
\newacronym{pwm}{PWM}{plane wave model} 
\newacronym{swm}{SWM}{spherical wave model} 
\newacronym{nlos}{NLOS}{non-line-of-sight} 
\newacronym[shortplural={MSS},firstplural={mobile satellite services (MSS)}]{mss}{MSS}{mobile satellite service}
\newacronym{lms}{LMS}{land mobile satellite} 
\newacronym{em}{EM}{electromagnetic} 
\newacronym{nato}{NATO}{North Atlantic Treaty Organization} 
\newacronym[shortplural={UHF},longplural={ultra high frequencies}]{uhf}{UHF}{ultra high frequency}
\newacronym{uca}{UCA}{uniform circular array} 
\newacronym{wss}{WSS}{wide sense stationary} 
\newacronym{pdp}{PDP}{power delay profile} 
\newacronym{rms}{RMS}{root mean square} 
\newacronym{vhf}{VHF}{very high frequency} 
\newacronym{cazac}{CAZAC}{constant amplitude auto-correlation} 
\newacronym{snr}{SNR}{signal to noise power ratio} 
\newacronym{cnr}{SNR}{signal to noise power ratio}
\newacronym{cir}{SIR}{signal to interference power ratio}
\newacronym{cinr}{SINR}{signal to interference and noise power ratio}  
\newacronym{sjr}{SJR}{signal to jammer power ratio} 
\newacronym{twta}{TWTA}{traveling wave tube amplifier} 
\newacronym{lnb}{LNB}{low-noise block-down-converter} 
\newacronym{MILSATCOM}{MILSATCOM}{military satellite communication} 
\newacronym{ECCM}{ECCM}{electronic counter counter measure} 
\newacronym[shortplural={EHF},longplural={extreme high frequencies}]{ehf}{EHF}{extreme high frequency}
\newacronym[shortplural={MPCs},longplural={multi path components}]{mpc}{MPC}{multi path component}
\newacronym{bpsk}{BPSK}{binary phase-shift keying} 
\newacronym{dbpsk}{DBPSK}{differential binary phase-shift keying} 
\newacronym{ccf}{CCF}{cross correlation function}
\newacronym{cfo}{CFO}{carrier frequency offset}
\newacronym{itu}{ITU}{International Telecommunication Union}
\newacronym{fec}{FEC}{forward error correction}
\newacronym{dpsk}{DPSK}{differential phase-shift keying}
\newacronym{cazac}{CAZAC}{constant amplitude zero auto-correlation}
\newacronym{cs}{CS}{chirp-sequence}
\newacronym{css}{CSS}{chirp-spread spectrum}
\newacronym{iot}{IoT}{Internet of Things}
\newacronym{m2m}{M2M}{machine-to-machine}
\newacronym{mtc}{MTC}{machine type communication}
\newacronym{mmtc}{mMTC}{massive machine type communication}
\newacronym{lpwan}{LPWAN}{low-power wide-area network}
\newacronym{ooc}{OOC}{optical orthogonal code}
\newacronym{mac}{MAC}{medium access control}
\newacronym{ra}{RA}{random access}
\newacronym{ssa}{SSA}{spread spectrum ALOHA}
\newacronym{sic}{SIC}{successive interference cancellation}
\newacronym{iic}{IIC}{iterative interference cancellation}
\newacronym{fer}{FER}{frame-error rate}
\newacronym{ucss}{UCSS}{Unipolar Coded Chirp-Spread Spectrum}
\newacronym{flr}{FLR}{frame loss ratio}
\newacronym{gnss}{GNSS}{global navigation satellite system}
\newacronym{gps}{GPS}{global positioning system}
\newacronym{prn}{PRN}{pseudo random number}
\newacronym{uw}{UW}{unique word}
\newacronym{nb}{NB}{narrow-band}
\newacronym{unb}{uNB}{ultra-narrow-band}

\newacronym{scada}{SCADA}{supervisory control and data acquisition}
\newacronym{noma}{NOMA}{non-orthogonal multiple access}
\newacronym{sdr}{SDR}{software defined radio}
\newacronym{ota}{OTA}{over-the-air}
\newacronym{buc}{BUC}{block up-converter}
\newacronym{muos}{MUOS}{Mobile User Objective System}
\newacronym{iobt}{IoBT}{Internet of Battlefield Things}
\newacronym{iort}{IoRT}{Internet of Remote Things}
\newacronym{wcdma}{WCDMA}{Wideband Code Division Multiple Access}
\newacronym{tdoa}{TDOA}{time difference of arrival}
\newacronym{fdoa}{FDOA}{frequency difference of arrival}
\newacronym{mse}{MSE}{mean squared error}
\newacronym{rmse}{RMSE}{root mean squared error}
\newacronym{crb}{CRLB}{Cramer Rao lower bound }
\newacronym{msb}{MSB}{most significant bit}
\newacronym{lsb}{LSB}{least significant bit}
\newacronym{ldro}{LDRO}{low data rate optimization}
\newacronym{lo}{LO}{local oscillator}
\newacronym{tcxo}{TCXO}{temperature compensated crystal oscillator}
\newacronym{crc}{CRC}{cyclic redundancy check}
\newacronym{psd}{PSD}{power spectral density}

%% file: figures/TikzDraw/overhead/source.tex

\tikzset{%
  block/.style    = {draw, thick, rectangle, minimum height = 3em,
      minimum width = 3em},
			 blockS/.style    = {draw, thick, rectangle, minimum height = 1em,
      minimum width = 1em},
  sum/.style      = {draw, circle, node distance = 2em,inner sep=1pt}, 
	 sumF/.style      = {draw, circle,fill, node distance = 2em,inner sep=1pt}, 
  add/.style      = {draw, circle, node distance = 2em,inner sep=10pt}, 
  input/.style    = {coordinate}, 
  output/.style   = {coordinate}, 
	textb/.style   = {inner sep=0pt,outer sep=0.5pt, minimum height = 1em,
      minimum width = 1em} ,
  dummy/.style   = {inner sep=0pt,outer sep=0.0pt} 
}
\newcommand*{\bitvector}[3]{
  \draw[fill=#3] (t_cur) -- ++( .05, .3) -- ++(#2-.1,0) -- ++(.05, -.3)
                         -- ++(-.05,-.3) -- ++(.1-#2,0) -- cycle;
  \path (t_cur) -- node[anchor=mid] {#1} ++(#2,0) node[time] (t_cur) {};
}
\newcommand*{\legendblock}[3]{
  \draw[fill=#3] (t_cur) -- ++( .05, .2) -- ++(#2-.1,0) -- ++(.05, -.2)
                         -- ++(-.05,-.2) -- ++(.1-#2,0) -- cycle;
  \path (t_cur) -- node[anchor=mid] {#1} ++(#2,0) node[time] (t_cur) {};
}

\newcommand*{\pause}[3]{
  \draw[fill=#3] (t_cur) -- ++( .01, .3) -- ++(#2-.02,0) -- ++(.01, -.3)
                         -- ++(-.01,-.3) -- ++(.02-#2,0) -- cycle;
  \path (t_cur) -- node[anchor=mid] {#1} ++(#2,0) node[time] (t_cur) {};
}

\begin{tikzpicture}[auto, thick, 
  node distance=0.1em, 
  >=triangle 45,
  every text node part/.style={align=center}
	]
  \tikzstyle{time}=[coordinate];

\node at (0,0) [dummy](t_cur){} ;

\draw[dotted] (t_cur) +(0,.25) node[above]  {.} -- ++(0,1) node[time](T0){} -- ++(0,0.5);
\bitvector{}{1.5}{black!30}; 
\node at (t_cur) [dummy,minimum width = 2em, right=0em of t_cur](t_cur){ ...  } ;
\node[dummy,minimum width = .1em, right=0em of t_cur](t_cur){} ;
\bitvector{}{1.5}{black!30}; 
\bitvector{}{0.375}{black!30}; 
 \draw[dotted] (t_cur) +(0,.25) node[above]  {.} -- ++(0,1) node[time](T1){} -- ++(0,0.5);
\bitvector{}{1.5}{white};
 \draw[dotted] (t_cur) +(0,.25) node[above]  {.} -- ++(0,1) node[time](T2){} -- ++(0,0.5);
\bitvector{}{1.5}{white};
\node at (t_cur) [dummy,minimum width = 2em, right=0em of t_cur](t_cur){ ...  } ;
\node[dummy,minimum width = .1em, right=0em of t_cur](t_cur){} ;
\bitvector{}{1.5}{white};
 \draw[dotted] (t_cur) +(0,.25) node[above]  {.} -- ++(0,1) node[time](T3){} -- ++(0,0.5);

\node at (0,1.9) [dummy](t_cur){} ;
\node [dummy, right=0em of t_cur](t1){\textbf{LoRa CSS packet structure}};
\draw[{Latex[length=1mm, width=2mm]}-{Latex[length=1mm, width=2mm]}] (T1) -- (T3)node [midway, above,xshift=5mm] {payload length $\Tpay$};
\draw[{Latex[length=1mm, width=2mm]}-{Latex[length=1mm, width=2mm]}] (T0) -- (T1)node [midway, above] {preamble length $\Tpre$};
\draw[{Latex[length=1mm, width=2mm]}-{Latex[length=1mm, width=2mm]}] ([yshift=-5mm] T1) -- ([yshift=-5mm] T2)node [midway, above] {$\TS$};

	%

	\node at (0,-1.0) [dummy](t_cur){} ;


\node [dummy, below = 2 of t_cur](t_cur){} ;
 \draw[dotted] (t_cur) +(0,.25) node[above]  {.} -- ++(0,1) node[time](T0){} -- ++(0,0.5);
\bitvector{ }{0.4}{black!30};
\bitvector{ }{0.4}{black!30};
\pause{}{0.05}{black!30};
\bitvector{ }{0.4}{black!30};
\bitvector{ }{0.4}{black!30};
\pause{}{0.02}{black!30};
\bitvector{ }{0.4}{black!30};
\bitvector{ }{0.4}{black!30};
\pause{}{0.07}{black!30};
 \draw[dotted] (t_cur) +(0,.25) node[above]  {.} -- ++(0,1) node[time](T1){} -- ++(0,0.5);
\bitvector{ }{0.4}{white};
\draw[dotted] (t_cur) +(0,.25) node[above] {.} -- ++(0,1)node[time] (T2) {.}-- ++(0,0.5);
\bitvector{ }{0.4}{white};
\pause{}{0.09}{black!30};
\bitvector{ }{0.4}{white};
\bitvector{ }{0.4}{white};
\pause{}{0.06}{black!30};
\bitvector{ }{0.4}{white};
\bitvector{ }{0.4}{white};
\pause{}{0.08}{black!30};
\bitvector{ }{0.4}{white};
\bitvector{ }{0.4}{white};
\pause{}{0.06}{black!30};
\bitvector{ }{0.4}{white};

\node[dummy,minimum width = 2em, right=0em of t_cur](t_cur){ ... } ;
\node[dummy,minimum width = .1em, right=0em of t_cur](t_cur){} ;

\pause{}{0.06}{black!30};
\bitvector{ }{0.4}{white};
\bitvector{ }{0.4}{white};
\draw[dotted] (t_cur) +(0,.25) node[above] {.} -- ++(0,1)node[time] (T3) {.}-- ++(0,0.5);

\draw[{Latex[length=1mm, width=2mm]}-{Latex[length=1mm, width=2mm]}] (T1) -- (T3)node [midway, above] {payload length $\Tpay$};
\draw[{Latex[length=1mm, width=2mm]}-{Latex[length=1mm, width=2mm]}] (T0) -- (T1)node [midway, above] {preamble len. $\Tpre$};
\draw[{Latex[length=1mm, width=2mm]}-{Latex[length=1mm, width=2mm]}] ([yshift=-5mm] T1) -- ([yshift=-5mm] T2)node [midway, above] {$\TS$};

\node at (0,-1.1) [dummy](t_cur){} ;
\node [dummy, right=0em of t_cur](t1){\textbf{UCSS packet structure}};



\end{tikzpicture}

%% file: tabledata.tex
\begin{tabular}{lll|l|l|l|l|l|l|llll|l|l|l|l|l|l|}
\cline{4-9} \cline{14-19}
                                                      &                                                   &               & \multicolumn{6}{l|}{\multirow{2}{*}{\textbf{LoRa}}}                                            &                                &                                           &                                                   &               & \multicolumn{6}{l|}{\multirow{2}{*}{\textbf{UCSS}}}                                               \\ \cline{1-3} \cline{11-13}
\multicolumn{1}{|l|}{\textbf{Parameter}}              & \multicolumn{1}{l|}{\textbf{Symbol}}              & \textbf{Unit} & \multicolumn{6}{l|}{}                                                                          & \multicolumn{1}{l|}{}          & \multicolumn{1}{l|}{\textbf{Parameter}}   & \multicolumn{1}{l|}{\textbf{Symbol}}              & \textbf{Unit} & \multicolumn{6}{l|}{}                                                                             \\ \cline{1-9} \cline{11-19} 
\multicolumn{1}{|l|}{\textbf{Setting Nr.}}            & \multicolumn{1}{l|}{\textbf{}}                    & \textbf{}     & \textbf{LS-1} & \textbf{LS-2}  & \textbf{LS-3} & \textbf{LS-4} & \textbf{LS-5} & \textbf{LS-6} & \multicolumn{1}{l|}{\textbf{}} & \multicolumn{1}{l|}{\textbf{Setting Nr.}} & \multicolumn{1}{l|}{\textbf{}}                    & \textbf{}     & \textbf{US-1} & \textbf{US-2} & \textbf{US-3} & \textbf{US-4} & \textbf{US-5}   & \textbf{US-6}   \\ \cline{1-9} \cline{11-19} 
\multicolumn{1}{|l|}{Sample Rate}                     & \multicolumn{1}{l|}{$\BW$}                        & kHz           & 62            & 20             & 20            & 20            & 20            & 20            & \multicolumn{1}{l|}{}          & \multicolumn{1}{l|}{Sample Rate}          & \multicolumn{1}{l|}{$\BW$}                        & kHz           & 62            & 20            & 20            & 20            & 20              & 20              \\ \cline{1-9} \cline{11-19} 
\multicolumn{1}{|l|}{Spreading}                       & \multicolumn{1}{l|}{$\loraSF$}                    &               & 9             & 9              & 10            & 11            & 12            & 12            & \multicolumn{1}{l|}{}          & \multicolumn{1}{l|}{Spreading}            & \multicolumn{1}{l|}{$\SF$}                        &               & 67            & 67            & 117           & 211           & 373             & 373             \\ \cline{1-9} \cline{11-19} 
\multicolumn{1}{|l|}{Code Rate}                       & \multicolumn{1}{l|}{$\loraCR$}                    &               & 0,5           & 0,5            & 0,5           & 0,5           & 0,5           & 1,0           & \multicolumn{1}{l|}{}          & \multicolumn{1}{l|}{Code Rate}            & \multicolumn{1}{l|}{$\CR$}                        &               & 0,5           & 0,5           & 0,5           & 0,5           & 0,5             & 0,5             \\ \cline{1-9} \cline{11-19} 
\multicolumn{1}{|l|}{LDR on}                          & \multicolumn{1}{l|}{$\loraLDRon$}                 &               & 0             & 1              & 1             & 1             & 1             & 1             & \multicolumn{1}{l|}{}          & \multicolumn{1}{l|}{PauseTimes}           & \multicolumn{1}{l|}{}                             & ms            & 72            & 224           & 224           & 224           & 224             & 61              \\ \cline{1-9} \cline{11-19} 
\multicolumn{1}{|l|}{User Bytes}                      & \multicolumn{1}{l|}{$\PL$}                        & bytes         & 16            & 8              & 8             & 8             & 8             & 4             & \multicolumn{1}{l|}{}          & \multicolumn{1}{l|}{User Bytes}           & \multicolumn{1}{l|}{$\PL$}                        & bytes         & 8             & 8             & 8             & 8             & 8               & 4               \\ \cline{1-9} \cline{11-19} 
\multicolumn{1}{|l|}{Symbols}                         & \multicolumn{1}{l|}{$\SmbNr$}                     &               & 40            & 24             & 24            & 24            & 16            & 12            & \multicolumn{1}{l|}{}          & \multicolumn{1}{l|}{Symbols}              & \multicolumn{1}{l|}{$\SmbNr$}                     &               & 134           & 134           & 134           & 134           & 134             & 70              \\ \cline{1-9} \cline{11-19} 
\multicolumn{1}{|l|}{Sym.   Duration}                 & \multicolumn{1}{l|}{$\TS$}                        & ms            & 8,3           & 25,6           & 51,2          & 102,4         & 204,8         & 204,8         & \multicolumn{1}{l|}{}          & \multicolumn{1}{l|}{Sym. Duration}        & \multicolumn{1}{l|}{$\TS$}                        & ms            & 1,1           & 3,4           & 5,9           & 10,6          & 18,7            & 18,7            \\ \cline{1-9} \cline{11-19} 
\multicolumn{1}{|l|}{Payl.   Duration}                & \multicolumn{1}{l|}{$\Tpay$}                      & ms            & 330           & 614            & 1229          & 2458          & 3277          & 2458          & \multicolumn{1}{l|}{}          & \multicolumn{1}{l|}{Payl. Duration}       & \multicolumn{1}{l|}{$\Tpay$}                      & ms            & 145           & 449           & 784           & 1414          & 2499            & 1306            \\ \cline{1-9} \cline{11-19} 
\multicolumn{1}{|l|}{Preamble Sym.}                   & \multicolumn{1}{l|}{$\PreSymbs$}                  &               & 12,25         & 12,25          & 12,25         & 12,25         & 12,25         & 12,25         & \multicolumn{1}{l|}{}          & \multicolumn{1}{l|}{Preamble Sym.}        & \multicolumn{1}{l|}{$\PreSymbs$}                  &               & 6             & 6             & 6             & 6             & 6               & 6               \\ \cline{1-9} \cline{11-19} 
\multicolumn{1}{|l|}{Prea.   Duration}                & \multicolumn{1}{l|}{$\Tpre$}                      & ms            & 101           & 314            & 627           & 1254          & 2509          & 2509          & \multicolumn{1}{l|}{}          & \multicolumn{1}{l|}{Prea. Duration}       & \multicolumn{1}{l|}{$\Tpre$}                      & ms            & 6             & 20            & 35            & 63            & 112             & 112             \\ \cline{1-9} \cline{11-19} 
\multicolumn{1}{|l|}{Packet Dur.}                     & \multicolumn{1}{l|}{$\Tpac$}                      & ms            & 431           & 928            & 1856          & 3712          & 5786          & 4966          & \multicolumn{1}{l|}{}          & \multicolumn{1}{l|}{Packet Dur.}          & \multicolumn{1}{l|}{$\Tpac$}                      & ms            & 224           & 693           & 1043          & 1701          & 2835            & 1479            \\ \cline{1-9} \cline{11-19} 
\multicolumn{1}{|l|}{Required SNR}                    & \multicolumn{1}{l|}{$\snrReq$}                    & dB            & -12,5         & -12,5          & -15,0         & -17,5         & -20,0         & -20,0         & \multicolumn{1}{l|}{}          & \multicolumn{1}{l|}{Required SNR}         & \multicolumn{1}{l|}{$\snrReq$}                    & dB            & -12,6         & -12,6         & -15,0         & -17,5         & -20,0           & -20,0           \\ \cline{1-9} \cline{11-19} 
\multicolumn{1}{|l|}{Data Rate}                       & \multicolumn{1}{l|}{$\DR$}                        & bit/s         & 296,7         & 69,0           & 34,5          & 17,2          & 11,1          & 6,4           & \multicolumn{1}{l|}{}          & \multicolumn{1}{l|}{Data Rate}            & \multicolumn{1}{l|}{$\DR$}                        & bit/s         & 286,1         & 92,3          & 61,3          & 37,6          & 22,6            & 21,6            \\ \cline{1-9} \cline{11-19} 
\multicolumn{1}{|l|}{\textbf{Overhead}}               & \multicolumn{1}{l|}{\textbf{$\OHLora$}}           & \textbf{}     & \textbf{23\%} & \textbf{34\%}  & \textbf{34\%} & \textbf{34\%} & \textbf{43\%} & \textbf{51\%} & \multicolumn{1}{l|}{\textbf{}} & \multicolumn{1}{l|}{\textbf{Overhead}}    & \multicolumn{1}{l|}{\textbf{$\OHUcss$}}           & \textbf{}     & \textbf{35\%} & \textbf{35\%} & \textbf{25\%} & \textbf{17\%} & \textbf{11,9\%} & \textbf{11,7\%} \\ \cline{1-9} \cline{11-19} 
\multicolumn{1}{|l|}{Max. CFO   Datasheet}            & \multicolumn{1}{l|}{$\FreqDriftMaxDS$}            & Hz            & 40,4          & 208,3          & 104,2         & 52,1          & 26,0          & 26,0          & \multicolumn{1}{l|}{}          & \multicolumn{1}{l|}{Max. CFO (13)}        & \multicolumn{1}{l|}{$\FreqDriftMaxDS$}            & Hz            & 463           & 149           & 85            & 47            & 27              & 27              \\ \cline{1-9} \cline{11-19} 
\multicolumn{1}{|l|}{\textbf{Max. Drift   Datasheet}} & \multicolumn{1}{l|}{\textbf{$\maxfreqDriftRate$}} & \textbf{Hz/s} & \textbf{93,5} & \textbf{224,5} & \textbf{56,1} & \textbf{14,0} & \textbf{4,5}  & \textbf{5,2}  & \multicolumn{1}{l|}{\textbf{}} & \multicolumn{1}{l|}{\textbf{Max. Drift}}  & \multicolumn{1}{l|}{\textbf{$\maxfreqDriftRate$}} & \textbf{Hz/s} & \textbf{3195} & \textbf{332}  & \textbf{109}  & \textbf{34}   & \textbf{11}     & \textbf{21}     \\ \cline{1-9} \cline{11-19} 
\end{tabular}

%% file: figures/pnProfiles.tex
%
%
\definecolor{mycolor1}{rgb}{0.00000,0.44700,0.74100}%
\definecolor{mycolor2}{rgb}{0.85000,0.32500,0.09800}%
\definecolor{mycolor3}{rgb}{0.92900,0.69400,0.12500}%
\definecolor{mycolor4}{rgb}{0.49400,0.18400,0.55600}%
\definecolor{mycolor5}{rgb}{0.46600,0.67400,0.18800}%
\definecolor{mycolor6}{rgb}{0.30100,0.74500,0.93300}%
\definecolor{mycolor7}{rgb}{0.63500,0.07800,0.18400}%
\begin{tikzpicture}

\begin{axis}[%
width=0.951\figurewidth,
height=\figureheight,
at={(0\figurewidth,0\figureheight)},
scale only axis,
xmode=log,
xmin=10,
xmax=10000,
xminorticks=true,
xlabel style={font=\color{white!15!black}},
xlabel={frequency offset in Hz},
ymin=-140,
ymax=0,
ylabel style={font=\color{white!15!black}},
ylabel={phase noise in dBc/Hz},
axis background/.style={fill=white},
xmajorgrids,
xminorgrids,
ymajorgrids,
legend style={legend cell align=left, align=left, draw=white!15!black},
ylabel near ticks, xlabel near ticks
]
\addplot [color=mycolor1, line width=1.0pt, mark=+, mark options={solid, mycolor1}]
  table[row sep=crcr]{%
10	-30\\
100	-100\\
1000	-120\\
10000	-130\\
};
\addlegendentry{PN profile 1}

\addplot [color=mycolor2, line width=1.0pt, mark=o, mark options={solid, mycolor2}]
  table[row sep=crcr]{%
10	-20\\
100	-90\\
1000	-120\\
10000	-130\\
};
\addlegendentry{PN profile 2}

\addplot [color=mycolor3, line width=1.0pt, mark=asterisk, mark options={solid, mycolor3}]
  table[row sep=crcr]{%
10	-15\\
100	-85\\
1000	-120\\
10000	-130\\
};
\addlegendentry{PN profile 3}

\addplot [color=mycolor4, line width=1.0pt, mark=triangle, mark options={solid, rotate=90, mycolor4}]
  table[row sep=crcr]{%
10	-10\\
100	-80\\
1000	-120\\
10000	-130\\
};
\addlegendentry{PN profile 4}

\addplot [color=mycolor5, line width=1.0pt, mark=square, mark options={solid, mycolor5}]
  table[row sep=crcr]{%
10	-70\\
100	-100\\
1000	-100\\
10000	-110\\
};
\addlegendentry{PN profile 5}

\addplot [color=mycolor6, line width=1.0pt, mark=x, mark options={solid, mycolor6}]
  table[row sep=crcr]{%
10	-70\\
100	-80\\
1000	-80\\
10000	-80\\
};
\addlegendentry{PN profile 6}

\addplot [color=mycolor7, line width=1.0pt, mark=triangle, mark options={solid, rotate=270, mycolor7}]
  table[row sep=crcr]{%
10	-60\\
100	-60\\
1000	-60\\
10000	-60\\
};
\addlegendentry{PN profile 7}

\end{axis}
\end{tikzpicture}%

%% file: figures/LoraPnSimFer.tex
%
%
\definecolor{mycolor1}{rgb}{0.00000,0.44700,0.74100}%
\definecolor{mycolor2}{rgb}{0.85000,0.32500,0.09800}%
\definecolor{mycolor3}{rgb}{0.92900,0.69400,0.12500}%
\definecolor{mycolor4}{rgb}{0.49400,0.18400,0.55600}%
\definecolor{mycolor5}{rgb}{0.46600,0.67400,0.18800}%
\definecolor{mycolor6}{rgb}{0.30100,0.74500,0.93300}%
\definecolor{mycolor7}{rgb}{0.63500,0.07800,0.18400}%
\begin{tikzpicture}

\begin{axis}[%
width=0.951\figurewidth,
height=\figureheight,
at={(0\figurewidth,0\figureheight)},
scale only axis,
xmin=-24,
xmax=-20,
xlabel style={font=\color{white!15!black}},
xlabel={SNR},
ymode=log,
ymin=0.001,
ymax=1,
yminorticks=true,
ylabel style={font=\color{white!15!black}},
ylabel={frame error rate},
axis background/.style={fill=white},
xmajorgrids,
ymajorgrids,
yminorgrids,
legend style={at={(0.03,0.03)}, anchor=south west, legend cell align=left, align=left, draw=white!15!black},
ylabel near ticks, xlabel near ticks
]
\addplot [color=mycolor1, dashed, line width=1.0pt, mark=+, mark options={solid, mycolor1}]
  table[row sep=crcr]{%
-24	0.980582524271845\\
-23	0.841666666666667\\
-22	0.442982456140351\\
-21	0.0742647058823529\\
-20	0.005\\
-19	0\\
-18	0\\
};
\addlegendentry{PN profile 1}

\addplot [color=mycolor2, dashed, line width=1.0pt, mark=o, mark options={solid, mycolor2}]
  table[row sep=crcr]{%
-24	0.990196078431373\\
-23	0.971153846153846\\
-22	0.651612903225806\\
-21	0.184643510054845\\
-20	0.025\\
-19	0.0005\\
-18	0\\
};
\addlegendentry{PN profile 2}

\addplot [color=mycolor3, dashed, line width=1.0pt, mark=asterisk, mark options={solid, mycolor3}]
  table[row sep=crcr]{%
-24	1\\
-23	1\\
-22	0.801587301587302\\
-21	0.412244897959184\\
-20	0.0911552346570397\\
-19	0.009\\
-18	0\\
};
\addlegendentry{PN profile 3}

\addplot [color=mycolor4, dashed, line width=1.0pt, mark=triangle, mark options={solid, rotate=90, mycolor4}]
  table[row sep=crcr]{%
-24	1\\
-23	1\\
-22	0.918181818181818\\
-21	0.573863636363636\\
-20	0.167774086378738\\
-19	0.019\\
-18	0\\
};
\addlegendentry{PN profile 4}

\addplot [color=mycolor5, dashed, line width=1.0pt, mark=square, mark options={solid, mycolor5}]
  table[row sep=crcr]{%
-24	0.990196078431373\\
-23	0.848739495798319\\
-22	0.385496183206107\\
-21	0.0737764791818846\\
-20	0.004\\
-19	0.0005\\
-18	0\\
};
\addlegendentry{PN profile 5}

\addplot [color=mycolor6, dashed, line width=1.0pt, mark=x, mark options={solid, mycolor6}]
  table[row sep=crcr]{%
-24	0.980582524271845\\
-23	0.885964912280702\\
-22	0.480952380952381\\
-21	0.066056245912361\\
-20	0.002\\
-19	0.0005\\
-18	0\\
};
\addlegendentry{PN profile 6}

\addplot [color=mycolor7, dashed, line width=1.0pt, mark=triangle, mark options={solid, rotate=270, mycolor7}]
  table[row sep=crcr]{%
-24	0.980582524271845\\
-23	0.885964912280702\\
-22	0.47196261682243\\
-21	0.0965583173996176\\
-20	0.004\\
-19	0\\
-18	0\\
};
\addlegendentry{PN profile 7}

\addplot [color=mycolor1, dashed, line width=1.0pt, mark=triangle, mark options={solid, mycolor1}]
  table[row sep=crcr]{%
-24	0.980582524271845\\
-23	0.795275590551181\\
-22	0.381132075471698\\
-21	0.08603066439523\\
-20	0.003\\
-19	0\\
-18	0\\
};
\addlegendentry{no phase noise}

\end{axis}
\end{tikzpicture}%

%% file: figures/UcssPnSimFer.tex
%
%
\definecolor{mycolor1}{rgb}{0.00000,0.44700,0.74100}%
\definecolor{mycolor2}{rgb}{0.85000,0.32500,0.09800}%
\definecolor{mycolor3}{rgb}{0.92900,0.69400,0.12500}%
\definecolor{mycolor4}{rgb}{0.49400,0.18400,0.55600}%
\definecolor{mycolor5}{rgb}{0.46600,0.67400,0.18800}%
\definecolor{mycolor6}{rgb}{0.30100,0.74500,0.93300}%
\definecolor{mycolor7}{rgb}{0.63500,0.07800,0.18400}%
\begin{tikzpicture}

\begin{axis}[%
width=0.951\figurewidth,
height=\figureheight,
at={(0\figurewidth,0\figureheight)},
scale only axis,
xmin=-24,
xmax=-20,
xlabel style={font=\color{white!15!black}},
xlabel={SNR},
ymode=log,
ymin=0.001,
ymax=1,
yminorticks=true,
ylabel style={font=\color{white!15!black}},
ylabel={frame error rate},
axis background/.style={fill=white},
xmajorgrids,
ymajorgrids,
yminorgrids,
legend style={at={(0.03,0.03)}, anchor=south west, legend cell align=left, align=left, draw=white!15!black},
ylabel near ticks, xlabel near ticks
]
\addplot [color=mycolor1, line width=1.0pt, mark=+, mark options={solid, mycolor1}]
  table[row sep=crcr]{%
-24	0.980582524271845\\
-23	0.918181818181818\\
-22	0.450892857142857\\
-21	0.0475\\
-20	0.0035\\
-19	0\\
-18	0\\
};
\addlegendentry{PN profile 1}

\addplot [color=mycolor2, line width=1.0pt, mark=o, mark options={solid, mycolor2}]
  table[row sep=crcr]{%
-24	1\\
-23	0.90990990990991\\
-22	0.457013574660633\\
-21	0.060226595110316\\
-20	0.004\\
-19	0\\
-18	0\\
};
\addlegendentry{PN profile 2}

\addplot [color=mycolor3, line width=1.0pt, mark=asterisk, mark options={solid, mycolor3}]
  table[row sep=crcr]{%
-24	1\\
-23	0.980582524271845\\
-22	0.664473684210526\\
-21	0.0974903474903475\\
-20	0.0055\\
-19	0\\
-18	0\\
};
\addlegendentry{PN profile 3}

\addplot [color=mycolor4, line width=1.0pt, mark=triangle, mark options={solid, rotate=90, mycolor4}]
  table[row sep=crcr]{%
-24	1\\
-23	1\\
-22	0.952830188679245\\
-21	0.627329192546584\\
-20	0.208247422680412\\
-19	0.0155\\
-18	0\\
};
\addlegendentry{PN profile 4}

\addplot [color=mycolor5, line width=1.0pt, mark=square, mark options={solid, mycolor5}]
  table[row sep=crcr]{%
-24	0.990196078431373\\
-23	0.918181818181818\\
-22	0.48792270531401\\
-21	0.0541264737406216\\
-20	0\\
-19	0.0005\\
-18	0\\
};
\addlegendentry{PN profile 5}

\addplot [color=mycolor6, line width=1.0pt, mark=x, mark options={solid, mycolor6}]
  table[row sep=crcr]{%
-24	1\\
-23	0.893805309734513\\
-22	0.417355371900826\\
-21	0.056965595036661\\
-20	0.0015\\
-19	0\\
-18	0\\
};
\addlegendentry{PN profile 6}

\addplot [color=mycolor7, line width=1.0pt, mark=triangle, mark options={solid, rotate=270, mycolor7}]
  table[row sep=crcr]{%
-24	1\\
-23	0.952830188679245\\
-22	0.5\\
-21	0.0670650730411687\\
-20	0.0015\\
-19	0\\
-18	0\\
};
\addlegendentry{PN profile 7}

\addplot [color=mycolor1, line width=1.0pt, mark=triangle, mark options={solid, mycolor1}]
  table[row sep=crcr]{%
-24	1\\
-23	0.901785714285714\\
-22	0.446902654867257\\
-21	0.044\\
-20	0.001\\
-19	0\\
-18	0\\
};
\addlegendentry{no phase noise}

\end{axis}
\end{tikzpicture}%

%% file: figures/driftSimSymErrs.tex
%
%
\begin{tikzpicture}[%
mark indices={2,10,20,30,40}
]

\begin{axis}[%
width=\figurewidth,
height=0.521\figureheight,
at={(0\figurewidth,0\figureheight)},
scale only axis,
xmode=log,
xmin=1,
xmax=7000,
xminorticks=true,
xlabel style={font=\color{white!15!black}},
xlabel={frequency drift rate $\freqDriftRate$ in Hz/s},
ymin=0,
ymax=15,
ylabel style={font=\color{white!15!black}},
ylabel={symbol errors},
axis background/.style={fill=white},
xmajorgrids,
xminorgrids,
ymajorgrids,
legend style={at={(0.5,1.03)}, anchor=south, legend cell align=left, align=left, draw=white!15!black},
legend columns=2, ylabel near ticks, xlabel near ticks
]
\addplot [color=blue, dashed, line width=1.0pt, mark=+, mark options={solid, blue}]
  table[row sep=crcr]{%
100	0\\
125	0\\
150	0\\
175	0\\
200	0\\
225	0\\
250	0\\
275	0\\
300	0\\
325	0.004\\
350	0.009\\
375	0.03\\
400	0.114\\
425	0.263\\
450	0.555\\
475	0.911\\
500	1.46\\
525	2.105\\
550	2.659\\
575	3.268\\
600	3.7572706935123\\
625	4.29790940766551\\
650	4.71177944862155\\
675	5.08253968253968\\
700	5.51231527093596\\
725	5.86528497409326\\
750	6.26136363636364\\
775	6.49152542372881\\
800	6.93478260869565\\
825	7.16296296296296\\
850	7.39855072463768\\
875	7.592\\
900	7.86614173228346\\
925	8.00787401574803\\
950	8.256\\
975	8.48760330578512\\
1000	8.62931034482759\\
1025	8.85344827586207\\
1050	8.9375\\
1075	9.27272727272727\\
1100	9.40178571428571\\
1125	9.57377049180328\\
1150	9.60344827586207\\
1175	9.7479674796748\\
1200	9.96774193548387\\
1225	10\\
1250	10.1333333333333\\
1275	10.2796610169492\\
1300	10.3697478991597\\
1325	10.4732142857143\\
1350	10.5762711864407\\
1375	10.5752212389381\\
1400	10.7647058823529\\
1425	10.7946428571429\\
1450	10.9245283018868\\
1475	11\\
1500	11.0654205607477\\
1525	11.1504424778761\\
1550	11.256880733945\\
1575	11.3942307692308\\
1600	11.4038461538462\\
};
\addlegendentry{LoRa LS-1, $\loraSF$ = 9}

\addplot [color=red, line width=1.0pt, mark=+, mark options={solid, red}]
  table[row sep=crcr]{%
2000	0\\
2105.26315789474	0\\
2210.52631578947	0\\
2315.78947368421	0\\
2421.05263157895	0\\
2526.31578947368	0.59\\
2631.57894736842	0.14\\
2736.84210526316	0.53\\
2842.10526315789	1.75\\
2947.36842105263	6.09\\
3052.63157894737	15.09\\
3157.8947368421	25.2447552447552\\
3263.15789473684	30.7117117117117\\
3368.42105263158	32.247619047619\\
3473.68421052632	35.2079207920792\\
3578.94736842105	34.3689320388349\\
3684.21052631579	34.3300970873786\\
3789.47368421053	34.8058252427184\\
3894.73684210526	32.6601941747573\\
4000	33.7476635514019\\
};
\addlegendentry{UCSS US-1, $\SF$ = 67}

\addplot [color=blue, dashed, line width=1.0pt, mark=o, mark options={solid, blue}]
  table[row sep=crcr]{%
50	0\\
55	0\\
60	0\\
65	0\\
70	0\\
75	0\\
80	0\\
85	0.001\\
90	0.014\\
95	0.26\\
100	1.148\\
105	2.175\\
110	3.122\\
115	4.09320695102686\\
120	4.88771929824561\\
125	5.71794871794872\\
130	6.37121212121212\\
135	6.93913043478261\\
140	7.58878504672897\\
145	8.09708737864078\\
150	8.64077669902913\\
155	9.1980198019802\\
160	9.53398058252427\\
165	10.1941747572816\\
170	10.5490196078431\\
175	10.8543689320388\\
180	11.1782178217822\\
185	11.5346534653465\\
190	11.9207920792079\\
195	12.1862745098039\\
200	12.6504854368932\\
};
\addlegendentry{LoRa LS-2, $\loraSF$ = 9}

\addplot [color=red, line width=1.0pt, mark=o, mark options={solid, red}]
  table[row sep=crcr]{%
100	0.02\\
115.789473684211	0.045\\
131.578947368421	0\\
147.368421052632	0.25\\
163.157894736842	0.21\\
178.947368421053	0\\
194.736842105263	0\\
210.526315789474	0\\
226.315789473684	0.025\\
242.105263157895	0.16\\
257.894736842105	0.125\\
273.684210526316	0.225\\
289.473684210526	0.74\\
305.263157894737	4.86\\
321.052631578947	19.3542857142857\\
336.842105263158	31.1517857142857\\
352.631578947368	33.2596153846154\\
368.421052631579	31.9433962264151\\
384.210526315789	35.2376237623762\\
400	34.5728155339806\\
};
\addlegendentry{UCSS US-2, $\SF$ = 67}

\addplot [color=blue, dashed, line width=1.0pt, mark=asterisk, mark options={solid, blue}]
  table[row sep=crcr]{%
0	0\\
5	0\\
10	0\\
15	0\\
20	0\\
25	0\\
30	0\\
35	0.063\\
40	1.687\\
45	3.27\\
50	4.52459016393443\\
55	5.64571428571429\\
60	6.45714285714286\\
65	7.14876033057851\\
70	7.93693693693694\\
75	8.24761904761905\\
80	8.85833333333333\\
85	9.1588785046729\\
90	9.69724770642202\\
95	9.98058252427184\\
100	10.125\\
};
\addlegendentry{LoRa LS-3, $\loraSF$ = 10}

\addplot [color=red, line width=1.0pt, mark=asterisk, mark options={solid, red}]
  table[row sep=crcr]{%
100	0\\
110.526315789474	0.03\\
121.052631578947	1.565\\
131.578947368421	20.4510869565217\\
142.105263157895	33.7433628318584\\
152.631578947368	36.5566037735849\\
163.157894736842	35.1121495327103\\
173.684210526316	36.9809523809524\\
184.210526315789	33.5701754385965\\
194.736842105263	33.027027027027\\
205.263157894737	35.2777777777778\\
215.789473684211	33.9339622641509\\
226.315789473684	33.9259259259259\\
236.842105263158	31.8738738738739\\
247.368421052632	33.2201834862385\\
257.894736842105	32.0566037735849\\
268.421052631579	30.9158878504673\\
278.947368421053	30.1456310679612\\
289.473684210526	30.0776699029126\\
300	31.9108910891089\\
};
\addlegendentry{UCSS US-3, $\SF$ = 113}

\addplot [color=blue, dashed, line width=1.0pt, mark=triangle, mark options={solid, rotate=90, blue}]
  table[row sep=crcr]{%
0	0\\
1.57894736842105	0\\
3.15789473684211	0\\
4.73684210526316	0\\
6.31578947368421	0\\
7.89473684210526	0\\
9.47368421052632	0.92\\
11.0526315789474	3.041\\
12.6315789473684	4.68170426065163\\
14.2105263157895	5.94623655913978\\
15.7894736842105	6.95070422535211\\
17.3684210526316	7.83760683760684\\
18.9473684210526	8.40869565217391\\
20.5263157894737	9\\
22.1052631578947	9.61739130434783\\
23.6842105263158	9.99152542372881\\
25.2631578947368	10.1578947368421\\
26.8421052631579	10.8099173553719\\
28.4210526315789	10.9910714285714\\
30	11.0267857142857\\
};
\addlegendentry{LoRa LS-4, $\loraSF$ = 11}

\addplot [color=red, line width=1.0pt, mark=triangle, mark options={solid, rotate=90, red}]
  table[row sep=crcr]{%
20	0\\
22.6315789473684	0\\
25.2631578947368	0\\
27.8947368421053	0\\
30.5263157894737	0\\
33.1578947368421	0\\
35.7894736842105	0\\
38.4210526315789	1.92\\
41.0526315789474	13.835\\
43.6842105263158	32.5348837209302\\
46.3157894736842	39.4571428571429\\
48.9473684210526	41.1442307692308\\
51.5789473684211	42.2912621359223\\
54.2105263157895	40.3177570093458\\
56.8421052631579	41.2115384615385\\
59.4736842105263	43.3592233009709\\
62.1052631578947	40.552380952381\\
64.7368421052632	41.0480769230769\\
67.3684210526316	43.0761904761905\\
70	41.7924528301887\\
};
\addlegendentry{UCSS US-4, $\SF$ = 211}

\addplot [color=blue, dashed, line width=1.0pt, mark=square, mark options={solid, blue}]
  table[row sep=crcr]{%
0	0\\
0.5	0\\
1	0\\
1.5	0\\
2	0\\
2.5	1.686\\
3	4.044\\
3.5	5.85714285714286\\
4	7.00729927007299\\
4.5	8.01680672268908\\
5	8.98305084745763\\
5.5	9.51724137931035\\
6	10.0088495575221\\
6.5	10.5045045045045\\
7	10.9917355371901\\
7.5	11.0238095238095\\
8	11.5546218487395\\
8.5	11.9652173913043\\
9	11.9909090909091\\
9.5	12.0090909090909\\
10	12.1651376146789\\
};
\addlegendentry{LoRa LS-5, $\loraSF$ = 12}

\addplot [color=red, line width=1.0pt, mark=square, mark options={solid, red}]
  table[row sep=crcr]{%
11	0\\
12.0526315789474	0\\
13.1052631578947	2.755\\
14.1578947368421	20.655\\
15.2105263157895	34.775\\
16.2631578947368	40.4038461538462\\
17.3157894736842	40.9903846153846\\
18.3684210526316	43\\
19.4210526315789	42.9029126213592\\
20.4736842105263	43.7087378640777\\
21.5263157894737	43.2427184466019\\
22.5789473684211	42.5339805825243\\
23.6315789473684	45.0095238095238\\
24.6842105263158	45.4285714285714\\
25.7368421052632	44.3238095238095\\
26.7894736842105	45.2815533980583\\
27.8421052631579	47.0990099009901\\
28.8947368421053	43.9411764705882\\
29.9473684210526	44.5728155339806\\
31	44.9108910891089\\
};
\addlegendentry{UCSS US-5, $\SF$ = 373}

\end{axis}
\end{tikzpicture}%

%% file: figures/driftSimFer.tex
%
%
\begin{tikzpicture}[%
mark indices={2,4,7}
]

\begin{axis}[%
width=\figurewidth,
height=0.514\figureheight,
at={(0\figurewidth,0\figureheight)},
scale only axis,
xmode=log,
xmin=1,
xmax=7000,
xminorticks=true,
xlabel style={font=\color{white!15!black}},
xlabel={frequency drift rate $\freqDriftRate$ in Hz/s},
ymode=log,
ymin=0.01,
ymax=1,
yminorticks=true,
ylabel style={font=\color{white!15!black}},
ylabel={frame error rate},
axis background/.style={fill=white},
xmajorgrids,
xminorgrids,
ymajorgrids,
yminorgrids,
legend style={at={(0.5,1.03)}, anchor=south, legend cell align=left, align=left, draw=white!15!black},
legend columns=2, ylabel near ticks, xlabel near ticks
]
\addplot [color=blue, dashed, line width=1.0pt, mark=+, mark options={solid, blue}]
  table[row sep=crcr]{%
100	0\\
125	0\\
150	0\\
175	0\\
200	0\\
225	0\\
250	0\\
275	0\\
300	0\\
325	0\\
350	0\\
375	0\\
400	0\\
425	0\\
450	0.001\\
475	0\\
500	0.002\\
525	0.009\\
550	0.023\\
575	0.048\\
600	0.112975391498881\\
625	0.17595818815331\\
650	0.2531328320802\\
675	0.320634920634921\\
700	0.497536945812808\\
725	0.523316062176166\\
750	0.573863636363636\\
775	0.570621468926554\\
800	0.731884057971015\\
825	0.748148148148148\\
850	0.731884057971015\\
875	0.808\\
900	0.795275590551181\\
925	0.795275590551181\\
950	0.808\\
975	0.834710743801653\\
1000	0.870689655172414\\
1025	0.870689655172414\\
1050	0.901785714285714\\
1075	0.834710743801653\\
1100	0.901785714285714\\
1125	0.827868852459016\\
1150	0.870689655172414\\
1175	0.821138211382114\\
1200	0.814516129032258\\
1225	0.827868852459016\\
1250	0.841666666666667\\
1275	0.855932203389831\\
1300	0.848739495798319\\
1325	0.901785714285714\\
1350	0.855932203389831\\
1375	0.893805309734513\\
1400	0.848739495798319\\
1425	0.901785714285714\\
1450	0.952830188679245\\
1475	0.901785714285714\\
1500	0.94392523364486\\
1525	0.893805309734513\\
1550	0.926605504587156\\
1575	0.971153846153846\\
1600	0.971153846153846\\
};
\addlegendentry{LoRa LS-1, $\loraSF$ = 9}

\addplot [color=red, line width=1.0pt, mark=+, mark options={solid, red}]
  table[row sep=crcr]{%
2000	0\\
2105.26315789474	0\\
2210.52631578947	0\\
2315.78947368421	0\\
2421.05263157895	0\\
2526.31578947368	0.02\\
2631.57894736842	0.005\\
2736.84210526316	0.015\\
2842.10526315789	0.06\\
2947.36842105263	0.18\\
3052.63157894737	0.455\\
3157.8947368421	0.706293706293706\\
3263.15789473684	0.90990990990991\\
3368.42105263158	0.961904761904762\\
3473.68421052632	1\\
3578.94736842105	0.980582524271845\\
3684.21052631579	0.980582524271845\\
3789.47368421053	0.980582524271845\\
3894.73684210526	0.980582524271845\\
4000	0.94392523364486\\
};
\addlegendentry{UCSS US1, $\SF$ = 67}

\addplot [color=blue, dashed, line width=1.0pt, mark=o, mark options={solid, blue}]
  table[row sep=crcr]{%
50	0\\
55	0\\
60	0\\
65	0\\
70	0\\
75	0\\
80	0\\
85	0\\
90	0\\
95	0\\
100	0.001\\
105	0.003\\
110	0.032\\
115	0.15955766192733\\
120	0.354385964912281\\
125	0.647435897435897\\
130	0.765151515151515\\
135	0.878260869565217\\
140	0.94392523364486\\
145	0.980582524271845\\
150	0.980582524271845\\
155	1\\
160	0.980582524271845\\
165	0.980582524271845\\
170	0.990196078431373\\
175	0.980582524271845\\
180	1\\
185	1\\
190	1\\
195	0.990196078431373\\
200	0.980582524271845\\
};
\addlegendentry{LoRa LS-2, $\loraSF$ = 9}

\addplot [color=red, line width=1.0pt, mark=o, mark options={solid, red}]
  table[row sep=crcr]{%
100	0.005\\
115.789473684211	0.005\\
131.578947368421	0\\
147.368421052632	0.01\\
163.157894736842	0.01\\
178.947368421053	0\\
194.736842105263	0\\
210.526315789474	0\\
226.315789473684	0.005\\
242.105263157895	0.005\\
257.894736842105	0.005\\
273.684210526316	0.01\\
289.473684210526	0.025\\
305.263157894737	0.14\\
321.052631578947	0.577142857142857\\
336.842105263158	0.901785714285714\\
352.631578947368	0.971153846153846\\
368.421052631579	0.952830188679245\\
384.210526315789	1\\
400	0.980582524271845\\
};
\addlegendentry{UCSS US1, $\SF$ = 67}

\addplot [color=blue, dashed, line width=1.0pt, mark=asterisk, mark options={solid, blue}]
  table[row sep=crcr]{%
0	0\\
5	0\\
10	0\\
15	0\\
20	0\\
25	0\\
30	0\\
35	0\\
40	0\\
45	0.025\\
50	0.236533957845433\\
55	0.577142857142857\\
60	0.721428571428571\\
65	0.834710743801653\\
70	0.90990990990991\\
75	0.961904761904762\\
80	0.841666666666667\\
85	0.94392523364486\\
90	0.926605504587156\\
95	0.980582524271845\\
100	0.901785714285714\\
};
\addlegendentry{LoRa LS-3, $\loraSF$ = 10}

\addplot [color=red, line width=1.0pt, mark=asterisk, mark options={solid, red}]
  table[row sep=crcr]{%
100	0\\
110.526315789474	0.005\\
121.052631578947	0.045\\
131.578947368421	0.548913043478261\\
142.105263157895	0.893805309734513\\
152.631578947368	0.952830188679245\\
163.157894736842	0.94392523364486\\
173.684210526316	0.961904761904762\\
184.210526315789	0.885964912280702\\
194.736842105263	0.90990990990991\\
205.263157894737	0.935185185185185\\
215.789473684211	0.952830188679245\\
226.315789473684	0.935185185185185\\
236.842105263158	0.90990990990991\\
247.368421052632	0.926605504587156\\
257.894736842105	0.952830188679245\\
268.421052631579	0.94392523364486\\
278.947368421053	0.980582524271845\\
289.473684210526	0.980582524271845\\
300	1\\
};
\addlegendentry{UCSS US1, $\SF$ = 113}

\addplot [color=blue, dashed, line width=1.0pt, mark=triangle, mark options={solid, rotate=90, blue}]
  table[row sep=crcr]{%
0	0\\
1.57894736842105	0\\
3.15789473684211	0\\
4.73684210526316	0\\
6.31578947368421	0\\
7.89473684210526	0\\
9.47368421052632	0\\
11.0526315789474	0.017\\
12.6315789473684	0.2531328320802\\
14.2105263157895	0.543010752688172\\
15.7894736842105	0.711267605633803\\
17.3684210526316	0.863247863247863\\
18.9473684210526	0.878260869565217\\
20.5263157894737	0.870689655172414\\
22.1052631578947	0.878260869565217\\
23.6842105263158	0.855932203389831\\
25.2631578947368	0.885964912280702\\
26.8421052631579	0.834710743801653\\
28.4210526315789	0.901785714285714\\
30	0.901785714285714\\
};
\addlegendentry{LoRa LS-4, $\loraSF$ = 11}

\addplot [color=red, line width=1.0pt, mark=triangle, mark options={solid, rotate=90, red}]
  table[row sep=crcr]{%
20	0\\
22.6315789473684	0\\
25.2631578947368	0\\
27.8947368421053	0\\
30.5263157894737	0\\
33.1578947368421	0\\
35.7894736842105	0\\
38.4210526315789	0.055\\
41.0526315789474	0.345\\
43.6842105263158	0.782945736434108\\
46.3157894736842	0.961904761904762\\
48.9473684210526	0.971153846153846\\
51.5789473684211	0.980582524271845\\
54.2105263157895	0.94392523364486\\
56.8421052631579	0.971153846153846\\
59.4736842105263	0.980582524271845\\
62.1052631578947	0.961904761904762\\
64.7368421052632	0.971153846153846\\
67.3684210526316	0.961904761904762\\
70	0.952830188679245\\
};
\addlegendentry{UCSS US1, $\SF$ = 211}

\addplot [color=blue, dashed, line width=1.0pt, mark=square, mark options={solid, blue}]
  table[row sep=crcr]{%
0	0\\
0.5	0\\
1	0\\
1.5	0\\
2	0\\
2.5	0\\
3	0.067\\
3.5	0.554945054945055\\
4	0.737226277372263\\
4.5	0.848739495798319\\
5	0.855932203389831\\
5.5	0.870689655172414\\
6	0.893805309734513\\
6.5	0.90990990990991\\
7	0.834710743801653\\
7.5	0.801587301587302\\
8	0.848739495798319\\
8.5	0.878260869565217\\
9	0.918181818181818\\
9.5	0.918181818181818\\
10	0.926605504587156\\
};
\addlegendentry{LoRa LS-5, $\loraSF$ = 12}

\addplot [color=red, line width=1.0pt, mark=square, mark options={solid, red}]
  table[row sep=crcr]{%
11	0\\
12.0526315789474	0\\
13.1052631578947	0.075\\
14.1578947368421	0.5\\
15.2105263157895	0.841666666666667\\
16.2631578947368	0.971153846153846\\
17.3157894736842	0.971153846153846\\
18.3684210526316	0.990196078431373\\
19.4210526315789	0.980582524271845\\
20.4736842105263	0.980582524271845\\
21.5263157894737	0.980582524271845\\
22.5789473684211	0.980582524271845\\
23.6315789473684	0.961904761904762\\
24.6842105263158	0.961904761904762\\
25.7368421052632	0.961904761904762\\
26.7894736842105	0.980582524271845\\
27.8421052631579	1\\
28.8947368421053	0.990196078431373\\
29.9473684210526	0.980582524271845\\
31	1\\
};
\addlegendentry{UCSS US1, $\SF$ = 373}

\end{axis}
\end{tikzpicture}%

%% file: figures/TikzDraw/measSetup/source.tex

\tikzset{%
  block/.style    = {draw, thick, rectangle, minimum height = 3em,
      minimum width = 3em},
			 blockS/.style    = {draw, thick, rectangle, minimum height = 1em,
      minimum width = 1em},
  sum/.style      = {draw, circle, node distance = 2em,inner sep=1pt}, 
  add/.style      = {draw, circle, node distance = 2em,inner sep=10pt}, 
	modu/.style      = {draw, circle, node distance = 2em,inner sep=10pt}, 
  input/.style    = {coordinate}, 
  output/.style   = {coordinate}, 
	textb/.style   = {inner sep=0pt,outer sep=0.5pt, minimum height = 1em,
      minimum width = 1em} ,
  dummy/.style   = {inner sep=0pt,outer sep=0.5pt} 
}

\newcommand{\BPF}[2] 
{  
\begin{scope}[transform shape,rotate=#2]
\draw[thick] (#1)node[](a){} +(-12pt,-12pt) rectangle +(12pt,12pt);
\draw (a) +(-8pt,0) to[bend left] +(0,0) edge[bend right] +(8pt,0);
\draw ([yshift=5pt]a) +(-8pt,0) to[bend left] +(0,0) to[bend right] +(8pt,0);
\draw ([yshift=-5pt]a) +(-8pt,0) to[bend left] +(0,0) edge[bend right] +(8pt,0);
\draw[rotate=20] ([yshift=5pt]a) +(-4pt,0) -- +(7pt,0);
\draw[rotate=20] ([yshift=-5pt]a) +(-7pt,0) -- +(4pt,0);
\end{scope}
}

\begin{tikzpicture}[auto, thick, 
  >=latex,
  every text node part/.style={align=center}
	]

\node[block]                      (tx) at (0,0) {DUT Tx\\at $\fTx$};
\node[dummy, right=5em of tx]     (mult)          {};
\node[block, right=4em of mult]   (bp)          {};
\node[block, right=2em of bp]     (rx)          {DUT Rx\\at $\fRx$};
\node[block, below=3em of mult]   (lo)          {Signal Generator\\ \vspace{0.3em} 
                                                \begin{tikzpicture}
                                                  \draw (0,0) --(0.3,0.3) -- (0.6,0) -- (0.9,0.3) -- (1.2,0) --(1.5,0.3);
                                                \end{tikzpicture}
                                                \\
                                                $f=\fRx-\fTx+\CFOtr(t)$
                                              };

\BPF{bp}{0};

\draw[thick]                          (mult) circle (8pt);
\draw[rotate=45, line width=0.5pt]    (mult)  +(0,-8pt) -- +(0,8pt);
\draw[rotate=-45,line width=0.5pt]    (mult)  +(0,-8pt) -- +(0,8pt);

\draw[<-] (mult)+(-8pt,0) -- (tx);
\draw[->] (mult)+(8pt,0)  -- (bp);
\draw[<-] (mult)+(0,-8pt) -- (lo);
\draw[->] (bp)            -- (rx);

\end{tikzpicture}

%% file: figures/combDriftMeas.tex
%
%
\begin{tikzpicture}

\begin{axis}[%
width=\figurewidth,
height=0.796\figureheight,
at={(0\figurewidth,0\figureheight)},
scale only axis,
unbounded coords=jump,
xmode=log,
xmin=1,
xmax=1000,
xminorticks=true,
xlabel style={font=\color{white!15!black}},
xlabel={frequency drift rate $\freqDriftRateMeas$ in Hz/s},
ymode=log,
ymin=0.01,
ymax=1,
yminorticks=true,
ylabel style={font=\color{white!15!black}},
ylabel={frame error rate},
axis background/.style={fill=white},
xmajorgrids,
xminorgrids,
ymajorgrids,
yminorgrids,
legend style={at={(0.5,1.03)}, anchor=south, legend cell align=left, align=left, draw=white!15!black},
legend columns=2, ylabel near ticks, xlabel near ticks
]
\addplot [color=blue, dashed, line width=1.0pt, mark=square, mark options={solid, blue}]
  table[row sep=crcr]{%
1	0\\
2	0.01\\
3.33333333333333	0.05\\
5	0.2\\
10	0.9\\
1000	nan\\
nan	nan\\
};
\addlegendentry{LoRa LS-5, $\loraSF$=12}

\addplot [color=blue, dashed, line width=1.0pt, mark=o, mark options={solid, blue}]
  table[row sep=crcr]{%
166.666666666667	0\\
200	0.013\\
222.222222222222	0.02\\
250	0.12\\
285.714285714286	0.25\\
333.333333333333	0.4\\
500	1\\
};
\addlegendentry{LoRa LS-2, $\loraSF$=9}

\addplot [color=red, line width=1.0pt, mark=square, mark options={solid, red}]
  table[row sep=crcr]{%
5	0\\
10	0.04\\
15	0.5\\
20	0.87\\
30	1\\
};
\addlegendentry{UCSS US-5, $\SF$=373}

\addplot [color=red, line width=1.0pt, mark=o, mark options={solid, red}]
  table[row sep=crcr]{%
300	0\\
500	0\\
550	0.1\\
600	0.46\\
700	1\\
};
\addlegendentry{UCSS US-2, $\SF$=67}

\end{axis}
\end{tikzpicture}%